# Measuring Hardware Impairments with Software-Defined Radios

Vuk Marojevic[*], Aditya V. Padaki[†], Raghunandan M. Rao[†], Jeffrey H. Reed[†]
[*]Dept. Electrical and Computer Engineering, Mississippi State University, Mississippi State, MS
[†]Wireless@VT, Bradley Dept. Electrical and Computer Engineering, Virginia Tech, Blacksburg, VA
{maroje|avpadaki|raghumr|reedjh}@vt.edu

*Abstract*—This Innovative Practice Full Paper introduces a novel tool for educating electrical engineering students about hardware impairments in wireless communications. A radio frequency (RF) front end is an essential part of a wireless transmitter or receiver. It features analog processing components and data converters which are driven by today's digital communication systems. Advancements in computing and software-defined radio (SDR) technology have enabled shaping waveforms in software and using experimental and easily accessible plug-and-play RF front ends for education, research and development. We use this same technology to teach nonlinear effects of RF front ends and their implications. It uses widely available RF instruments and components and SDR technology—well-established affordable hardware and free open source software—to teach students how to characterize the nonlinearity of RF receivers while providing hands-on experience with SDR tools. We present the hardware, software and procedures of our laboratory session that enable easy reproducibility in other classrooms. We discuss different forms of evaluating the suitability of the new class modules and conclude that it provides a valuable learning experience that bolsters the theory that is typically provided in lectures only.

*Keywords—software-defined radio, radio frequency receiver, nonlinearity, hands-on learning.*

## I. INTRODUCTION

This Innovative Practice Full Paper introduces a novel tool for educating electrical engineering students about hardware impairments in wireless communications. Students and researchers often neglect the fact that a radio frequency (RF) system is not ideal. An RF front end is an essential part of a wireless transmitter or receiver. It enables wireless communications and features analog processing components and data converters which are driven by today's digital communication systems. Advancements in computing and software-defined radio (SDR) technology facilitate defining the radio waveform in software and using experimental and easily accessible plug-and-play RF front ends for education, research and development. We use this same technology to teach nonlinear effects of RF front ends and their implications. Our focus is on radio receivers, although nonlinear RF problems occur at the transmitter as well and deserve their own experiments.

Inspired by research in the field of dynamic spectrum access and the implications of RF front end nonlinearities, we developed laboratory sessions to provide a valuable hands-on experience to students on some of the important hardware impairments of wireless communications receivers. The students learn to (1) set up a testbed and define a test procedure that allows reproducible measurements, (2) operate SDRs, and (3) compare theoretical with practical results and analyze the deviations. As opposed to related hand-on exercises, this one is challenging because it uses experimental RF hardware, which is far from ideal for showing how theory matches practice. The advantage of using such hardware is that students learn about the inherent difficulty of real-life measurements when dealing with real devices and their characterization. This is an invaluable experience in the advent of the Internet of Things.

The sessions that we developed consist of hardware, software, laboratory instructions and assignments. Only commercial off-the shelf hardware components are used, including Ettus Research Universal Software Radio Peripheral (USRP), a spectrum analyzer, RF cables, combiners and attenuators and computers. The software is open source and was developed using GNU Radio Companion and distributed to the students through a virtual machine image (native installation is also possible) to avoid hardware-software compatibility problems since students are encouraged to use their own computers.

The *RF hardware impairments laboratory* was launched in fall 2016 at Virginia Tech in the graduate Software Radios class and occupied two sessions of 3.5 hours each, with an additional session to allow students to complete the exercises and assignments. This paper introduces the sessions, discusses the logistical and technical challenges and evaluates the intervention using different mechanisms. More precisely, we use a series of quizzes and homework assignments that were scheduled before, during and after the sessions. The rest of the paper is organized as follows: Section II provides the necessary background and discusses some of the related work. Section III described the objectives, methodology and tools. Section IV introduces our software laboratories, which are evaluated in Section V. Section VI concludes the paper.

## II. BACKGROUND AND RELATED WORK

RF front ends contain several analog components that are inherently nonlinear. These include the high-power amplifier (HPA) and the low-noise amplifier (LNA) at the transmitter and receiver, respectively, mixers, and so forth. Receiver nonlinearity has been studied by several researchers and the polynomial approximation model is widely used to describe this nonlinearity [1] [2]. The 3$^{rd}$ order intercept point is commonly used to characterize the nonlinearity of a receiver. It can be obtained as illustrated in Fig. 1 using empirical data. The two-tone test, which is later described, is a way of obtaining such data. The first order or fundamental output has a linear input-output power relationship. The third order output has a slope of three and occurs because of receiver nonlinearity. More precisely, 3$^{rd}$ order intermodulation products appear at the frequencies $2f_1 - f_2$ and $2f_2 - f_1$, where $f_1$ and $f_2$ are the frequencies of the two fundamental tones that are fed into the RF device under test.

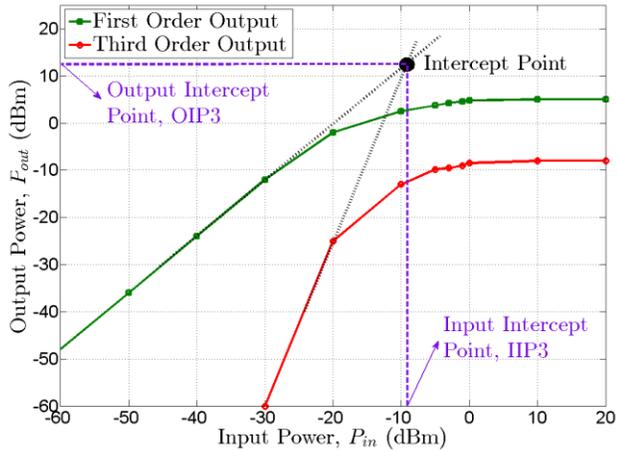

Fig. 1. Third order intercept point.

While the theory of 3$^{rd}$ order products is well understood, it has gained attention in the context of spectrum sharing because of the heterogeneous RF devices operating in adjacent bands at different power levels. With a wideband preselection filter, which facilitates access to different channels and poor nonlinearity characteristics, a high power adjacent channel signal can block the signal of interest even when the receiver is not driven into saturation [3]. This is known as the weak nonlinearity region.

As wireless network evolve and 5G networks are currently being standardized, testing becomes critical for those systems to perform efficiently and coexist with other systems [4]. Since this paper is about educating on RF impairments using SDRs, we focus our literature review on related contributions to education and those that use similar tools or principles.

Helaly and Adnani [5] present an SDR instrumentation system to analyze RF signals and systems as an alternative to using traditional RF instruments. Guzelgoz and Arlslan [6] introduce a wireless communications systems laboratory course that provides a set of analysis tools for simulations and practical analyses of wireless systems. Tripathi et al. [7] propose an SDR solution to improve the quality of 4G signals. More precisely, they implement a digital predistortion to counter the distortion introduced by the HPA and produce a cleaner transmit signal. Digital predistortion is an effective and important technique for high peak-to-average power ration waveforms such as orthogonal frequency division multiplexing that is employed in 4G long-term evolution (LTE) and 5G new radio (NR) [8]. Martinek, et al. [9] use SDRs to create software tools and instruments that can model and visualize different wireless channel effects and RF impairments. SDRs can be used to build real-time channel emulators [10] that can be used to create controlled radio environments for research and education. SDRs also facilitate the acquisition and recording of RF signals and their playback [11]. This technique is convenient for feeding real-world signals into wireless testbeds [12], which are popular for research, development and education.

## III. OBJECTIVE, METHODOLOGY, AND TOOLS

### A. Objective

The goal is to experience RF nonlinearity of RF receivers. The students therefore measure the nonlinear characteristic of an RF receiver while gaining hands-on experience with SDRs.

### B. Methodology

The students get access to software and hardware and can use their own laptops to set up an SDR system, configure the hardware and perform the measurements in a controlled environment. More precisely, RF cables are used as opposed to antennas. Portability is ensured by offering virtual machine images that can be played with the VMware Player, which is available for free download from the VMware Web Site.

### C. Tools

The tools used in these laboratories are SDR hardware and software, a spectrum analyzer (SA), RF cables, combiners and attenuators. Using GNU Radio Companion, the students can build their own radios using readily available building blocks. Here they develop simple transmitter and receiver flow graphs for RF signal generation, acquisition and spectrum visualization. The SA allows for calibration of the uncalibrated SDR system. The RF components are used to set up the test system that allows reproducible experiments. We provide virtual machine (VM) images so that students can use their own computers to implement the SDR system. Fig. 2 shows a screenshot with one USRP connected to a computer and accessed from a VM.

## IV. LABORATORY

We introduce two laboratory sessions of two to three hours each. Lab-1 allows the students to get familiarized with the hardware and software and providing the essential tools for measuring and quantifying the non-linear characteristic of the RF front end.

### A. Lab-1: USRP Power Calibration

This laboratory session introduces the USRP and provides hands-on experience. The objective is to calibrate the USRP power level for a specific frequency in preparation of the characterization of the USRP RF front end.

In this laboratory the students learn how to use a USRP to generate and capture signals and calibrate the signal power. The components used in this laboratory are shown in Table I. Fig. 3. depicts the experimental setup and Fig. 4 the transmitter and receiver waveforms or GNU radio flow graphs used in Lab-1. These flow graphs do not need to be provided, but this depends on the instructor and the prior experience of the students. Note that the blocks have many configurable radio and processing parameters. For example, the sample rate or FFT size of the Sink receiver waveform can be downscaled if the CPU is the bottleneck, which is often the case when using personal laptops. These flow graphs illustrate how easy it is to build the SDR transmitter and receiver for the desired experiments. Not shown is the amount of tools that come with GNU Radio and its graphical user interface GNU Radio Companion to build more complex communications systems simply by dragging and connecting readily available processing blocks.

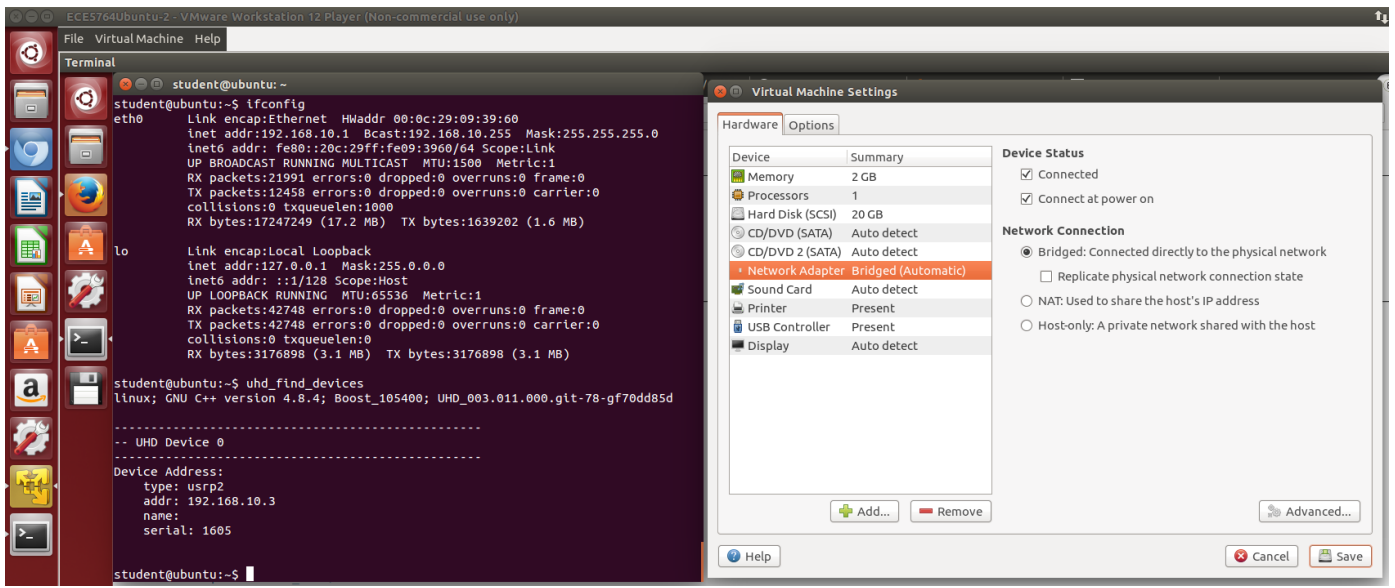

Fig. 2. USRP connected to a computer and accessed from a VM.

The procedure is as follows:

*1) System setup:* The students check out the components and set up the measurement system as illustrated in Fig. 3.

*2) Transmit USRP A – Tone Generation:* Using GNU Radio Companion the students build a flowgraph that generates a tone sent to USRP A.

*3) Receive USRP B:* Using GNU Radio Companion the students build a flowgraph that captures RF signals with USRP B and plots the FFT.

*4) Power level calibration:* With the help of the SA, the students calibrate the power level for different TX gains of USRP A and 0 dB Rx Gain of USRP B.

TABLE I. LAB-1 EQUIPMENT LIST.

| Component | Model or Characteristic | Quantity |
|---|---|---|
| USRP | B210 | 2 |
| RF cables | With SMA connectors | 1 |
| Attenuators | Fixed attenuators 2-30 dB | |
| Spectrum Analyzer | Tektronix SA 2500 | 1 |
| Computers | With USB 3 ports | 2 |
| Software | Ubuntu with GNU Radio Companion or VMware Player and provided image | |

### B. Lab-2: Two-Tone Test and IIP3 Measurement

This laboratory session introduces the two-tone test using a USRP to characterize another USRP's RF front end. The objective is to generate 3rd order intermodulation products to empirically obtain the third-order intercept point of the device under test (DUT).

In this session the students learn how to use a USRP to generate two tones using two RF chains, combine the signals through a RF power combiner, and empirically evaluate the third-order Intermodulation Intercept Point (IIP3) of a USRP. Table II shows the equipment list. We leverage the two channels of the B210 to generate the two tones. Fig. 5 indicates the setup and Fig. 6 the GNU Radio flow graph of the transmitter. The receiver is identical to Lab-1.

The procedure is as follows:

*1) System setup:* The students check out the components and set up the measurement system as illustrated in Fig. 5.

*2) Transmit USRP A – Tone Generation:* Using GNU Radio Companion the students build a flowgraph that generates two sinosoidal signal sources sent to USRP A, leveraging the two Tx/Rx chains of the B210.

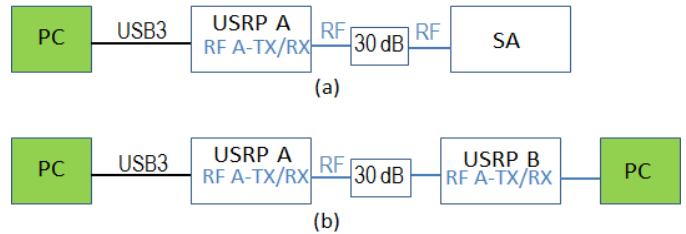

Fig. 3. Experimental setup for Lab-1 with spectrum analyzer (a) and SDR receiver (b).

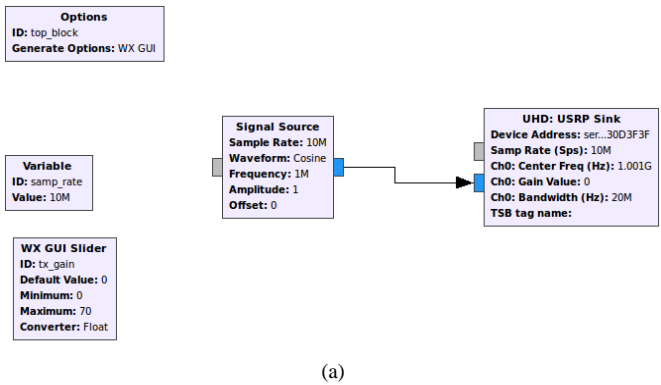

(a)

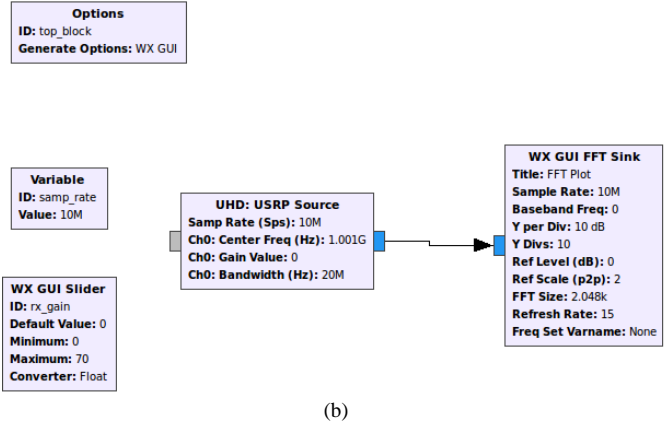

(b)

Fig. 4. GNU radio flow graphs defining the SDR transmitter (a) and receiver (b) of Lab-1.

TABLE II. LAB-2 EQUIPMENT LIST.

| Component | Model or Characteristic | Quantity |
|---|---|---|
| USRP | B210 | 2 |
| RF cables | With SMA connectors | 3 |
| Attenuators | Fixed attenuators 2-30 dB | |
| 3-way RF power combiner | Mini-Circuits | 1 |
| 50 Ohm matched termination | Mini-Circuits | 1 |
| Spectrum Analyzer | Tektronix SA 2500 | 1 |
| Computers | With USB 3 ports | 2 |
| Software | Ubuntu with GNU Radio Companion or VMware Player and provided image | |

*3) Receive USRP B:* Using GNU Radio Companion the students build a flowgraph that captures RF signals with USRP B and plots the FFT. Fig. 7 shows an illustrative capture of two fundamental tones and their 3$^{rd}$ order intermodulation products.

*4) Power level calibration:* With the help of the SA, the students calibrate the power level for different Tx gains of USRP A and 0 dB Rx Gain of USRP B.

*5) IIP3 measurements:* For a 70 dB Rx Gain of USRP B, the students find the Tx gain that shows third order intermods without saturating the transmitter (USRP A). They increase the Tx gain in 2 dB steps and note down the magnitude of the fundamental tones and third order products observed with the FFT Sink block on PC2, which is connected to USRP B. The students take at least 10 data points and plot the calibrated data points to empirically find the 3rd order intercept point (IP3) of USRP B. The IP3 is found at the intersection of the extended fundamental and 3rd order curves, as previously illustrated in Fig. 1.

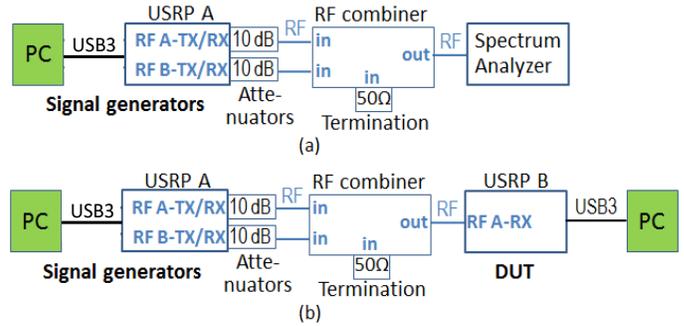

Fig. 5. Experimental setup for Lab-2 with spectrum analyzer (a) and SDR receiver (b).

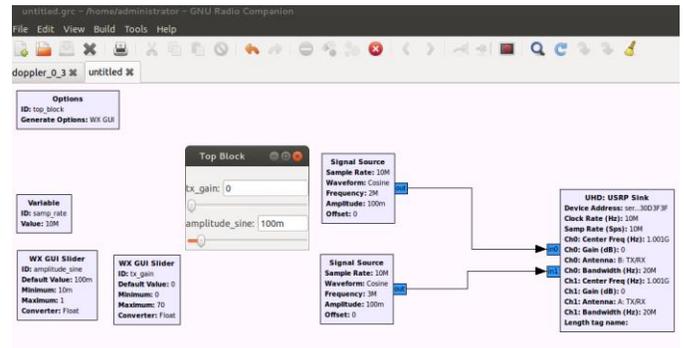

Fig. 6. GNU radio flow graphs defining the SDR transmitter of Lab-2.

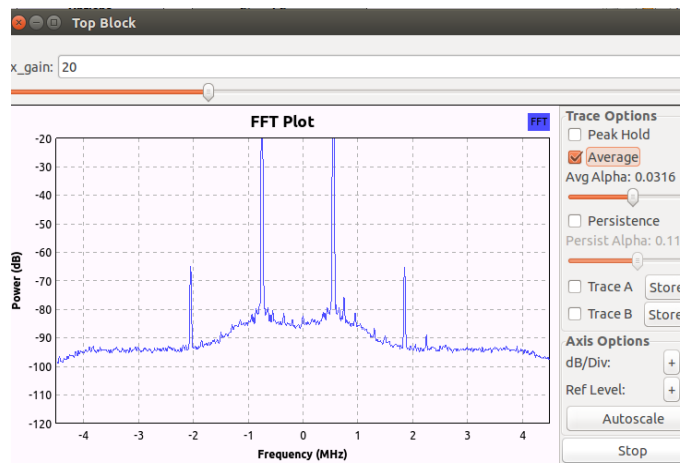

Fig. 7. Fundamental tones and 3$^{rd}$ order intermodulation products at the receiver as observed by the FFT Sink module. This setup uses a center frequency of $f_0$ = 900.75 MHz and fundamental tones are at 900 and 901.3 MHz, or -0.75 and 0.55 MHz relative to $f_0$ (0 MHz in the FFT Plot). The 3$^{rd}$ order intermodulation products are at 2*900 − 901.3 = 898.7 MHz or -2.05 MHz on the relative scale and 2*901.3 − 900 = 902.6 MHz or 1.85 MHz.

## V. EVALUATION

We apply several mechanisms to engage students in this hands-on learning opportunity and provide continuous evaluation and feedback. We use these mechanisms to assess the suitability of the laboratory.

### A. Student Evaluation

Students in this class are exposed to lectures, SDR laboratories, quizzes, and projects. The focus here are the new laboratories as described in Section III, but they need to put into context. After receiving lectures on RF related subjects, the students participate in hands-on software laboratory sessions. Before Lab-1, we give a short quiz for us to evaluate the general understanding of the material that was previously covered in class. Then, we provide a quick recap of the theory and how the students will use it to characterize receiver nonlinearity in the laboratories. This is followed by Lab-1. Lab-2 is scheduled for another day. The schedule of the laboratory sessions is flexible. We recommend not scheduling them back to back to give the students some time to assimilate the material.

A second quiz is scheduled at the beginning of the second laboratory session. Both quizzes are not previously announced. The second quiz asks fundamental and practical question related to the subject of matter and is similar, but not identical to the first.

We recommend at least 3 hours for each session and offering at least one additional open laboratory session, where student groups can redo or verify their results, as needed. Along with the laboratory material and instructions, the students get a worksheet with questions. The first set of questions are directly related to the experiments. The second set of questions requires additional research. Responses to the first set are to be delivered as part of the laboratory report and responses to the second set of questions as a homework deliverable. In total, we have two laboratory reports and two homework assignments related to these software laboratories.

### B. Laboratory Evaluation

The laboratory has been designed for the SDR class at Virginia Tech and had 15 participants. We use different laboratory assessment mechanisms. One is based on grading. We look at the evolution of grades before and after the laboratories and compare equivalent activities, such as pre and post-laboratory quizzes. The outcomes for the two quizzes show a positive student learning experience. The first quiz had low scores, whereas the second had medium to high scores on average. The two laboratory reports and the two related homework deliverables show a similar tendency, from medium to high scores. Whereas more data is needed for statistical analysis of scores, we conclude that this hands-on intervention is engaging students and motivating them to study and master the subject. Questions are defined in such a way that obliges the students to study the theory and match the experiments with fundamental principles in RF. The students also get exposed to using modern SDR hardware and software tools as well as RF components and equipment. As opposed to traditional SDR software sessions, these laboratories also look at the internals of SDR hardware peripherals as opposed to treating them as black boxes.

The laboratory instructor and two graduate students evaluated student progress and discussed their findings. Our experience is that most students have actively participated in the laboratories and contributed to the group success. Several groups took advantage of the open laboratory session to validate or finish their work. We base our observations on student participation during the official laboratory sessions and the open laboratory sessions, their engagement and creativity, the questions they asked, and so forth.

A third way we use to evaluate the success or failure of the learning method looks at student retention. Out of the 15 students in the class, six joined our research group and became graduate research assistants participating in various research projects. These students have performed very well in class and the laboratories and have done outstanding work since then. (Note that more students are involved in our group, but here we discuss on the new students that joined our group after attending the class.) Those six students have been working on SDR projects and use SDR hardware and software for their research.

Two undergraduate students from two distinct US universities have joined our group during the summers of 2017 and 2018 as part of the National Science Foundation's Research Experience for Undergraduates (REU) program. These students used these laboratory sessions to get experience with SDRs and learn how to analyze and characterize receiver RF nonlinearity in the first week of their research projects with minimal support.

## VI. CONCLUSIONS

This paper presents our SDR laboratory sessions that educate about RF receiver nonlinearity and how to empirically characterize it. This is motivated by the fact that theoretical concepts are best absorbed by students when they experience the practical effects and implications. We leveraged our research on the practical implications of heterogeneous RF characteristics of radios that share spectrum to teach well-established RF concepts using modern an accessible SDR technology. We introduced two laboratories that use SDRs to measure the $3^{rd}$ order intercept point using traditional techniques. The students get exposed to SDR hardware and software. Our approach is unique in that it provides invaluable hands-on experience with SDRs in combination with exposure to traditional RF issues, which are often neglected in research. Our experience is that this method helps bolstering theoretical principles and accomplishing the multidisciplinary learning objectives of a software radio engineering class. All the laboratory material, including instructions, software and VM images, are available for free download and enable reproducing these sessions in other classes, at others schools, or for self-learning purposes.

ACKNOWLEDGMENT

This work was in part supported by the National Science Foundation (NSF) under Grant CNS-1564148.